\documentclass[conference]{IEEEtran}
\IEEEoverridecommandlockouts
\usepackage{cite}
\usepackage{amsmath,amssymb,amsfonts}
\usepackage{algorithmic}
\usepackage{graphicx}
\usepackage{textcomp}
\usepackage{xcolor}
\usepackage{amsmath}
\usepackage{array}

\usepackage{graphicx}
\usepackage{algorithm}
\usepackage{algorithmic}
\usepackage{booktabs}
\usepackage{multirow}
\usepackage{amsfonts}
\usepackage{amssymb}

\usepackage{nicefrac}
\usepackage{wrapfig}
\usepackage{bm}

\makeatletter
\newcommand{\ssymbol}[1]{}
\makeatother

\def\BibTeX{{\rm B\kern-.05em{\sc i\kern-.025em b}\kern-.08em
    T\kern-.1667em\lower.7ex\hbox{E}\kern-.125emX}}
\begin{document}

\title{A Secure and Efficient Data Deduplication Scheme with Dynamic Ownership Management in Cloud Computing}
\author{
\IEEEauthorblockN{Xuewei Ma, Wenyuan Yang, Yuesheng Zhu, Zhiqiang Bai}
\IEEEauthorblockA{\textit{School of Eletronic and Computer Engineering, Peking University} \\
mxw971201@stu.pku.edu.cn, wyyang@pku.edu.cn, zhuys@pku.edu.cn, baizq@pku.edu.cn}
}

\maketitle

\begin{abstract}
Encrypted data deduplication is an important technique for saving storage space and network bandwidth, which has been widely used in cloud storage. Recently, a number of schemes that solve the problem of data deduplication with dynamic ownership management have been proposed. However, these schemes suffer from low efficiency when the dynamic ownership changes a lot. To this end, in this paper, we propose a novel server-side deduplication scheme for encrypted data in a hybrid cloud architecture, where a public cloud (Pub-CSP) manages the storage and a private cloud (Pri-CSP) plays a role as the data owner to perform deduplication and dynamic ownership management. Further, to reduce the communication overhead we use an initial uploader check mechanism to ensure only the first uploader needs to perform encryption, and adopt an access control technique that verifies the validity of the data users before they download data. Our security analysis and performance evaluation demonstrate that our proposed  server-side deduplication scheme has better performance in terms of security, effectiveness, and practicability compared with previous schemes. Meanwhile, our method can efficiently resist collusion attacks and duplicate faking attacks. 
\end{abstract}


\section{Introduction}
Nowadays, with the rapid growth of data volumes, there are urged demands for secure places where private data can be safely stored. Outsourcing the big data to the cloud is an efficient way to solve this problem \cite{Ibrahim2015The,2021Security}. Despite all the advantages of cloud computing \cite{Duan2016Cloud}, duplicated data still waste abundant storage space and network bandwidth, and make data management more complicated \cite{2017Heterogeneous}. 

The deduplication \cite{2002Venti} is a process that identifies the same data by data similarity, which allows the cloud storage provider save the storage by storing only a single copy of the data owned by multiple owners. However, the existing schemes related to deduplication still incur some problems about dynamic ownership management and access control.

First of all, to protect privacy, many users encrypt data before uploading it to the cloud storage. Since the encryption key is randomly generated, the same data encrypted with different keys will produce different ciphertext, which will hinder deduplication. To solve this problem, some deduplication schemes propose that the owners of the same file share the same encryption key \cite{1984Security,2013DupLESS,2013Anonymous,2014Secure,2015A,2015Interactive,2017Attribute,2020Lightweight,2021An}. However, most of them do not consider the dynamic ownership changes that happen frequently in cloud storage service \cite{Hur2016Secure}. The cloud users should be revoked from the valid ownership list once they request the cloud storage provider for data deletion/modification.

Second, to address dynamic ownership management, many schemes \cite{Hur2016Secure,2016Secure,2017Heterogeneous,Shen2019Multi,2018DedupDUM,2021Enhanced} were proposed by using either trusted third party or semi-trusted third party to do proxy re-encryption work, such as Authority Party ($AP$) or The Public Cloud Provider ($Pub-CSP$). On the one hand, it could be difficult to implement a trusted third party in practical applications \cite{2020SecDedup}. On the other hand, some schemes can not resist collusion attacks when the third party colludes with unauthorized users.

In this paper, we propose an novel scheme, which aims at efficiently solving the problem of deduplication with frequent cloud user revocation and new cloud user joining in cloud computing.
In particular, different from existing data deduplication methods, which employ either trusted/semi-trusted third party to do proxy re-encryption work, our proposed scheme designs a hybrid cloud architecture, which includes a public cloud and further introduces a private cloud. 
In implementations, the introduced private cloud in our scheme is involved as a data owner and a proxy at the same time to 1) control access to outsourced data by performing re-encryption techniques and 2) manage the dynamic ownership when real data owner is offline or revoke his/her ownership.

Furthermore, we propose to enhance our scheme in terms of efficiency by 1) ensuring that the data owner performs encryption only when he/she is the initial uploader; 2) presenting an access control technique that verifies the validity of the data users before they download data; and 3) requiring the public cloud server can send ciphertext to the cloud user only when the cloud users are in the ownership list.
Therefore, the abundant communication cost will be reduced.
We evaluate the performance of our scheme through security analysis, comparison with existing work, and implementation-based performance evaluation.

Overall, our contributions are as follows:

\begin{itemize}
    \item We propose a novel scheme, which introduces a hybrid cloud architecture including a public cloud and a private cloud, to efficiently solve the problem of deduplication with frequent cloud user revocation and new cloud user joining in cloud computing.
    
    \item We further propose to improve our scheme in efficiency. The extensive experiments prove the security, effectiveness, and efficiency of our method, which can significantly outperform previous data deduplication methods.
\end{itemize}

\section{RELATED WORK}

\subsection{Deduplication without Dynamic Ownership Management}

To solve deduplication problem, convergent encryption \cite{2002Reclaiming} was proposed, in which a data user got the key of data $F$ by computing its hash code $K=H(F)$. Since data $F$ will be encrypted with $K$, whoever holds the same data can produce the exact same encrypted data.

A server-aided encryption scheme (DupLESS) for data deduplication was proposed by Bellare et al. \cite{2013DupLESS}. In DupLESS, an independent key server generates the key, which suffers from large computation time in block-level deduplication since it takes long time to generate keys. Liu et al. \cite{2015Secure} proposed a client-side encryption that requests the data owner to do ownership check and deduplication, which is impractical. Cui et al. \cite{2017Attribute} proposed a deduplication scheme by using attribute-based access control technique under hybrid cloud environment.

Recently, a scheme \cite{2020SecDedup} performed duplication check by short hash. Although it can resist offline brute-force, the collision rate will be relatively high because different data may have the same short hash. Later, the advanced scheme \cite{2021PCSP} used light weight techniques. However, this scheme does not apply to the situation where there is massive duplicated data, and is more suitable for individual users to store data on cloud disks. The drawback of these schemes is not considering the dynamic ownership management among the data users. 

\subsection{Deduplication with Dynamic Ownership Management}
To address dynamic ownership management, many schemes were proposed by using either a trusted third party or semi-trusted third party to do proxy re-encryption work, such as AP or Pub-CSP. Wen et al.\cite{2016Secure} constructed a convergent key sharing scheme. But this work requires the data user to encrypt/decrypt convergent keys and recover them from secret shares, which is unrealistic since the data user’s computation power is limited. A server-side deduplication scheme was proposed by Hur et al. \cite{Hur2016Secure} in which the new cloud user joining was not considered since the total number of data users was fixed. Later, an enhanced scheme proposed by Yan et al. \cite{2017Heterogeneous} provided a heterogeneous data storage management. However, it is not practical for the owner to be always online and send personalized keys to the holders. Beside, once the owner is offline, the access control will be entirely dependent on AP. Premkamal et al. \cite{2021Enhanced} proposed an enhanced scheme with attribute-based access control by using the group key with the help of trusted entities.

Although these schemes address dynamic ownership management, there still are some security flaws. Since most of the schemes used either trusted/semi-trusted third party to do proxy re-encryption work, some data owners do not like to authorize a third party to control their data. Besides, taking scheme \cite{2017Heterogeneous} as an example, once the malicious user conspires with the third party (such as a proxy server), it will happen that unauthorized users $u$ who only has access to file $A$ can also have invalid access to file $B$. This is the problem we intend to solve in this study.

	\begin{figure}[t]
	    \centering
	    \includegraphics[width=1\linewidth]{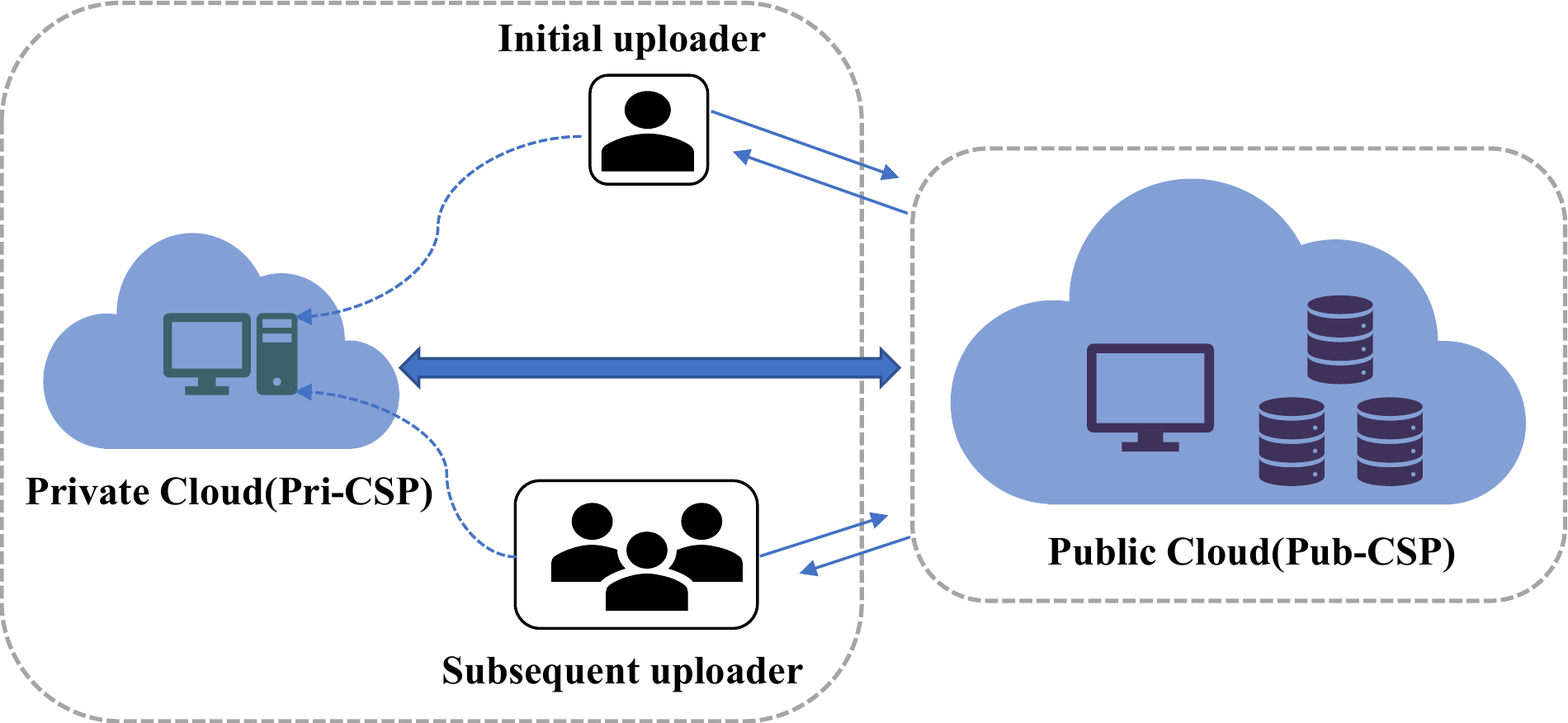}
	    \caption{Architecture of a data deduplication system.}
	\end{figure}

\section{SYSTEM MODEL}
In this section, we describe the data deduplication system and define the adversary model. Only the file-level deduplication is considered in this paper.

\subsection{Hybrid Architecture for Secure Deduplication}
Figure 1 shows the architecture of the data deduplication system, which consists of three entities stated as follows.

\begin{itemize}
	\item
	\textbf{Data users (DU).} This is an entity that wants to outsource data to Pub-CSP and access data later. In the authorized deduplication system, each user is issued a pair of secret key and public key $(sk_{u}, pk_{u})$ about Proxy Re-Encryption (PRE) and $(SK_{u},PK_{u})$ for signature. Moreover, if a data user is the first one to upload file $F_{A}$, she/he will be defined as the owner $u_{1, A}$; if the file $F_{A}$ already exist, the user will be defined as the holder $u_{i, A}$. 
	
	\item
	\textbf{Public Cloud (Pub-CSP).} The entity that provides a data storage service in the public cloud. In this paper, Pub-CSP is assumed to be always online and capable to provide abundant storage.
	
	\item
	\textbf{Private Cloud (Pri-CSP).} The entity that provides an execution environment and infrastructure for data users as an interface between DU and Pub-CSP, since the computing resources of DU are limited and Pub-CSP is not fully trusted in practice. Pri-CSP maintains an ownership list for storing data, which is composed of a hash code set for the stored data, identities, and re-encrypted keys belonging to the owners.
\end{itemize}

In this system, we assume that the infrastructure of Pri-CSP is exclusively used by a single organization consisting of multiple users. It is owned, managed, and operated by the organization itself. Therefore, Pri-CSP can be trusted by all entities. Pub-CSP will strictly follow system design and protocols strictly, while they may still be curious about the raw data of data users. Moreover, We assume that Pri-CSP and DU would never collude with Pub-CSP due to different business interests.

\subsection{Security Requirements}
We follow \cite{Hur2016Secure} to ensure the following security requirements: Data privacy, Data consistency, Ownership verification, Ownership revocation, Collusion resistance.

	\noindent
	\textbf{Data privacy.} Protecting the raw data from the Pub-CSP server and the unauthorized data users.

	\noindent
	\textbf{Data consistency.} Ensuring tag consistency against any poison attacks. Any modifications of the ciphertext can be verified by the authorized data users.

	\noindent 
	\textbf{Ownership verification.} Any unauthorized data users who can not be verified his/her ownerships should be prevented from accessing to the ciphertext or any messages related to decryption stored in the cloud.

	\noindent 
	\textbf{Ownership revocation.} 
	Any authorized data users who request the cloud storage provider for deletion/modification of the data $F$ should be revoked from the valid ownership list and accessing to the data $F$ anymore.

	\noindent 
	\textbf{Collusion resistance.} Any unauthorized data users who do not have valid ownership of data should unable to access the raw data even if they collude with each other or collude with Pub-CSP.

\section{PRELIMINARIES AND DEFINITION}
\subsection{Proxy Re-Encryption (PRE) }

Proxy re-encryption is initially proposed by Blaze et al. \cite{1998Divertible} called as atomic proxy functions, in which data users can decrypt ciphertext encrypted by key $A$ with another key $B$ without leaking any message about encryption/decryption key or plaintext. In this article, we have three actors in PRE:

\begin{itemize}
\item
\textbf{Data Owner}: authorizes decryption rights to data user;

\item
\textbf{Data User}: receives decryption rights to access the encrypted data;

\item
\textbf{Proxy}: performs re-encryption functions to allow Data User to access the encrypted data;

\end{itemize}

Specifically, the Data Owner $A$ encrypts data $F$ by his/her public key, $C_{A} = En(pk_{A}, F)$. If the Data User $B$ wants to decrypt the ciphertext $C_{A}$, he/she needs to get decryption rights from Data Owner $A$ with his/her public key $pk_{B}$. If $A$ agrees, he/she will produce an authorized key by $rk_{A\to B} = RG(sk_{A},pk_{B})$ and send it to the proxy.

In our scheme, Pri-CSP plays a role as \textbf{Data Owner} and \textbf{Proxy} at the same time to reduce client-side computation overhead. Table 1 summarizes the notations used in this paper.

\subsection{Defination}
In this section, we define a secure deduplication system for encrypted data with dynamic ownership management. The details of algorithms used in our scheme will be shown as follows. 

\noindent\textit{1) System Setup}

\begin{itemize}
\item 
$\boldsymbol{KeyGen1(u_{id})}$. Based on the system parameters, user $u$ generates its own key pair $(sk_{u}, pk_{u})$ about PRE and $(SK_{u}, PK_{u})$ for signature by inputting the unique $u_{id}$. Meanwhile, Pri-CSP gets  $(sk_{u_{0}}, pk_{u_{0}})$ by having $u_{0,id}$ as owner $u_{0}$.

\end{itemize}

\noindent\textit{2) Data Encryption and Decryption}
\begin{itemize}
\item
$\boldsymbol{Encrypt(K, F)}$. For plaintext data $F$, data owner $u_{1}$ encrypts it with the symmetric key $K$ to get ciphertext $CT$.

\item
$\boldsymbol{Decrypt(K, CT)}$. Data holder $u_{i}$ decrypts $CT$ with key $K$ and outputs $F$.
\end{itemize}

\noindent\textit{3) Key Control based on PRE Operated by Pri-CSP}

\begin{itemize}
\item 
$\boldsymbol{KeyGen2(\Lambda)}$ takes a security parameter as input, and outputs a random symmetric key $K$.

\item 
$\boldsymbol{En(pk_{u_{0}}, K)}$ takes $pk_{u_{0}}$ and the symmetric key $K$ as input and outputs an encrypted key $EK$.

\item 
$\boldsymbol{RG(sk_{u_{0}}, pk_{u_{i}})}$ outputs re-encryption key $rk_{u_{0}\to u_{i}}$ by taking $sk_{u_{0}}$ and $pk_{u_{i}}$ as input. 

\item 
$\boldsymbol{ReEn(rk_{u_{0}\to u_{i}}, EK)}$ takes input $rk_{u_{0} \to u_{i}}$ and $EK$, and outputs $REK_{u_{i}} = En(pk_{u_{i}},K)$ that can be decrypted with $sk_{u_{i}}$ by $De(sk_{u_{i}}, REK_{u_{i}})$.

\end{itemize}

\begin{table}[t]
\centering
\caption{Notation.}
\resizebox{0.45\textwidth}{!}{
\begin{tabular}{|l|l|}
\hline
\textbf{Notations} & \textbf{Description}                     \\ \hline
$F_{id}/F$         & The duplicated data $F$                    \\ \hline
$u_{id}/u$         & The users of cloud service               \\ \hline
$K/K'$             & The symmetric key of $F$/ Updated key      \\ \hline
$EK/EK'$           & The encrypted $K/K'$               \\ \hline
$REK/REK'$         & The encrypted $EK/EK'$ by PRE              \\ \hline
$sk_{u_{0}}$       & The secret key of the Pri-CSP about PRE  \\ \hline
$pk_{u_{0}}$       & The public key of the Pri-CSP about PRE  \\ \hline
$sk_{u}$           & The secret key of $u$ about PRE            \\ \hline
$pk_{u}$           & The public key of $u$ about PRE            \\ \hline
$CT/CT'$           & The encrypted data $F$/ Renewed ciphertext \\ \hline
$H(*)$             & The hash function                        \\ \hline
$HC(F)$            & The hash code set of data $F$              \\ \hline
\end{tabular}
}
\end{table}

\section{PROPOSED DEDUPLICATION SCHEME}

\subsection{Data Deduplication}
Figure 2 shows the procedure of initial upload by owner $u_{1}$. Figure 3 shows the procedure of deduplication when subsequent uploaders $u_{2}$ uploads the same data with $u_{1}$. We assume that Pri-CSP plays a role as owner $u_{0}$ to control dedupliacation for owner $u_{1}$.

\begin{itemize}

\item \smallskip\textbf{Step 1 - Key Generation:} After system parameter generation, each DU asks $KeyGen1(u_{id})$ to generate key pair $(sk_{u}, pk_{u})$ for PRE and $(SK_{u}, PK_{u})$ for signature. Meanwhile, Pri-CSP gets $(sk_{u_{0}}, pk_{u_{0}})$ by $KeyGen1(u_{0})$.

\item \smallskip\textbf{Step 2 - Duplication Check:} DU $u_{1}$ stores data $F$ at Pub-CSP. He/She calculates $H(F)$, signs it with $SK_{u_{1}}$ and sends data package $dp=\left\{ H(F), sign(H(F), SK_{u_{1}})\right\}$ to Pub-CSP. The duplication check will be performed by Pub-CSP to verify if the same data has been stored already after verifying the signature. If the check is positive, go to Step \textbf{5}. Otherwise, go to Step \textbf{3} to request key for encryption.

\item \smallskip\textbf{Step 3 - Data Storage:} When Pub-CSP defines $u_{1}$ is the first uploader of data $F$, it generates a unique $F_{id}$ for $F$ and contacts Pri-CSP to get key for encryption. 
Pri-CSP generates a random symmetric key $K$, and encryps it with $pk_{u_{0}}$ to get $EK$, then applys PRE to $EK$ to get cipherkey $REK_{u_{1}}$ for DU $u_{1}$. Pub-CSP sends $(F_{id}, REK_{u_{1}})$ to $u_{1}$ after receiving it from Pri-CSP. 
User $u_{1}$ decrypts $REK_{u_{1}}$ by  $De(sk_{u_{1}}, REK_{u_{1}})$ with his/her own private key to get $K$, and then $u_{1}$ encrypts data $F$ by $Encrypt(K, F)$. 
Moreover, user $u_{1}$ will randomly select several indexes: $X = \left\{x_{1},x_{2},\dots,x_{n}\right\}$ that indicate the specific parts of $F$ (e.g., $x_{1}$ equals first 0.5\% of $F$; $x_{2}$ equals first 2.5\% of $F$). Then, based on the index $X$, $u_{1}$ calculates the hash code set of data $F$ as $HC(F) = \left\{HC(F_{1}),HC(F_{2}),\dots,HC(F_{N})\right\}$. 
Then $u_{1}$ sends the data package $dp_1 = \left\{u_{1,id},REK_{u_{1}},CT\right\}$ to Pub-CSP and sends the data package $dp_2 = \left\{F_{id},u_{1,id},X,HC(F)\right\}$ to Pri-CSP. Last but not least, Pub-CSP and Pri-CSP both maintain the ownership list seperately for each data, what saved in Pub-CSP is $P_1$, and what saved in Pri-CSP is $P_2$, where $X'$ and $HC'(F)$ is randomly selected from $X$ and $HC(F)$.
\begin{equation}
\begin{aligned}
    &P_1 =\{F_{id}, CT,H(F), (u_{1,id}, REK_{u_{1}})\}  \\
    &P_2 = \left\{F_{id}, X', HC'(F), EK, (u_{1,id}, rk_{u_{0}\to u_{1}})\right\} 
\end{aligned}
\end{equation}

\item \smallskip\textbf{Step 4 - Duplicated Data Upload:} Later on, DU $u_{2}$ wants to store the same data $F$ at Pub-CSP by sending the data package $dp = \left\{H(F),sign(H(F),SK_{u_{2}})\right\}$.

\begin{figure}[t]
    \centering
    \includegraphics[width=1\linewidth]{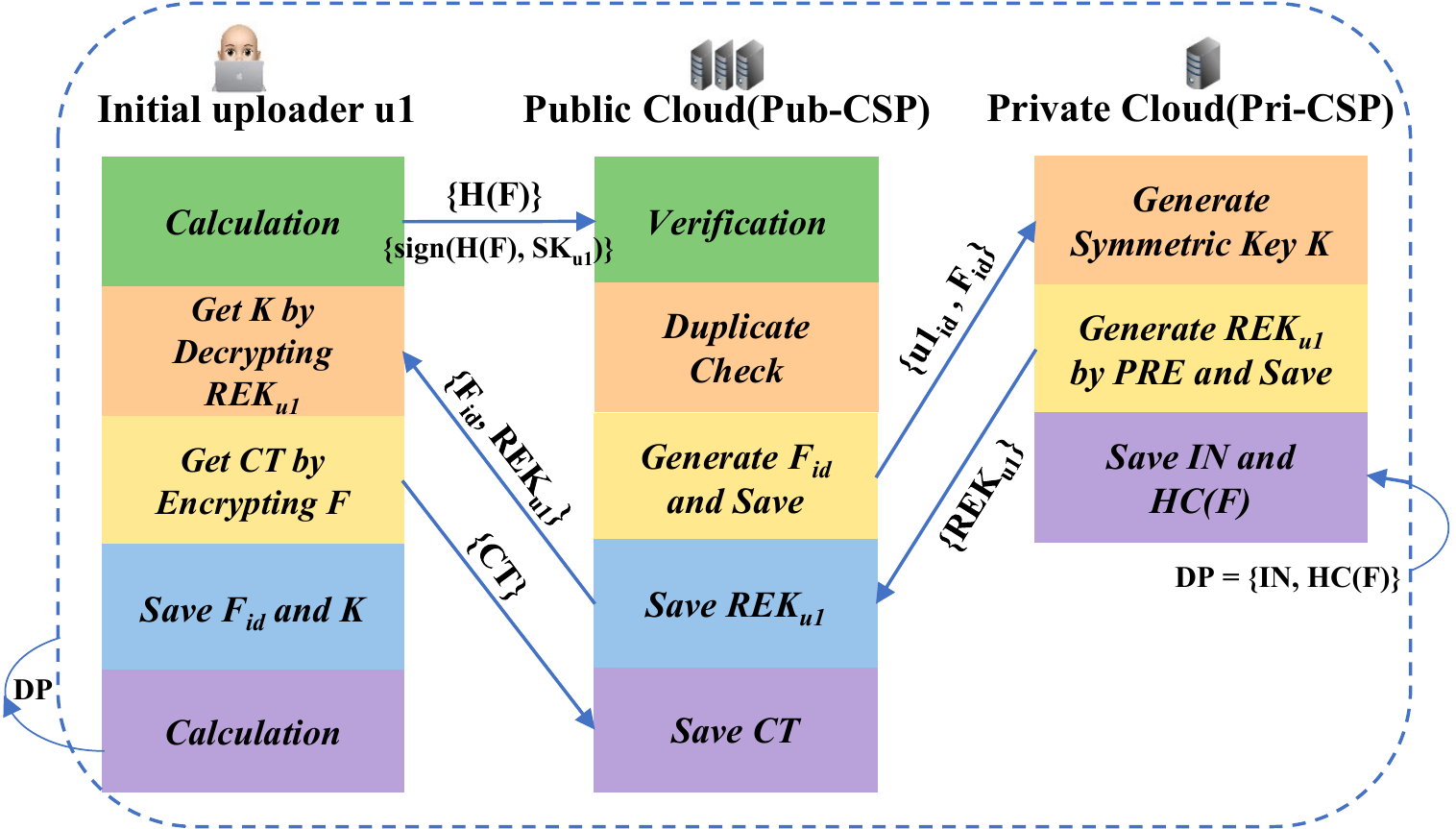}
    \caption{Procedure of initial upload by owner $u_{1}$.}
\end{figure}

\item \smallskip\textbf{Step 5 - Deduplication:} Pub-CSP performs duplication check as in step \textbf{2} after verifying that $u_{2_{id}}$ is not in the ownership list. When the duplication check is positive, Pub-CSP contacts Pri-CSP for deduplication. Pri-CSP further verifies the ownership of DU $u_{2}$ by challenging the hash code set of $F$ before performing deduplication, which ensures the data ownership, since there is a probability that $H(F)$ is eavesdropped or gained by the malicious party. If the ownership verification is positive, Pri-CSP generates the re-encrypted key $REK_{u_{2}}$ and sends it back to Pub-CSP for saving. Once Pub-CSP receives $REK_{u_{2}}$, it sends $REK_{u_{2}}$ with ciphertext $CT$ to $u_{2}$. When $u_{2}$ gets $REK_{u_{2}}$, he/she decrypts it with $sk_{u_{2}}$ to get $K$. Then $u_{2}$ gets plaintext of data $F$ by running $Decrypt(K, CT)$ and checks data consistency, while if $u_{2}$ does not want to check data consistency immediately, he/she can check it by downloading $CT$ and $H(F)$ later. Through data deduplication, both $u_{1}$ and $u_{2}$ can access the same data $F$ that is stored only once at Pub-CSP.

\end{itemize}

\begin{figure}[t]
    \centering
    \includegraphics[width=1\linewidth]{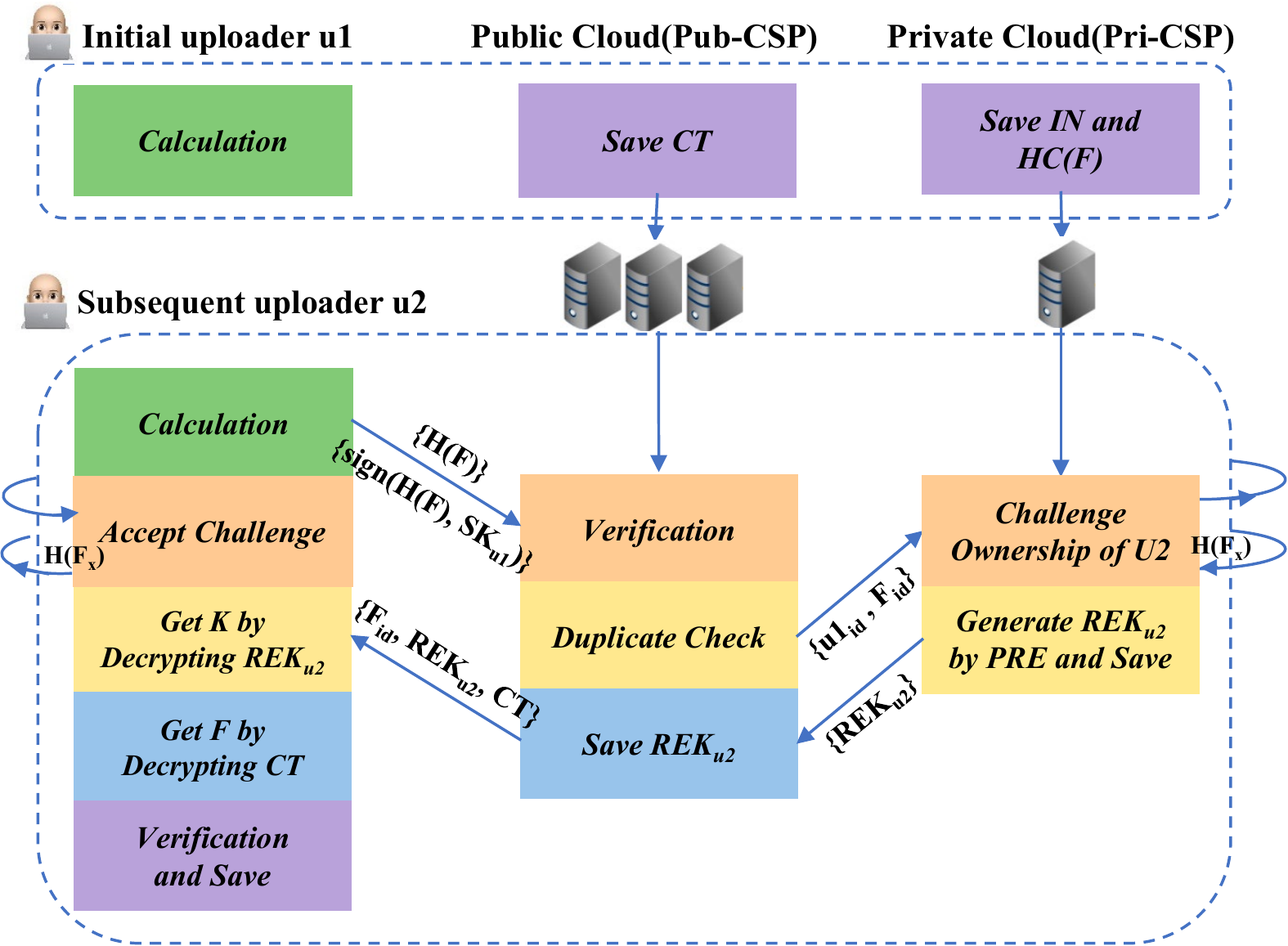}
    \caption{Procedure of deduplication.}
\end{figure}

\subsection{Ownership Revocation}
Any authorized data users who request Pub-CSP for the deletion/modification of their data stored in the cloud storage should be removed from the valid ownership list, and be prevented from accessing the previously saved data.

\smallskip\noindent\textbf{Ownership Revocation of Data Holder}. Figure 4 shows the procedure of data update by DU $u_{2}$ in the context of data deduplication.

\begin{itemize}

\item \smallskip\textbf{Step 1:} User $u_{2}$ sends a request of data deletion to Pub-CSP by providing $dp = \left\{F_{id}, sign(F_{id}, SK_{u_{2}})\right\}$.

\item \smallskip\textbf{Step 2:} Pub-CSP deletes the storage record of $u_{2}$ after verifying the signature. Then Pub-CSP asks owner $u_{1}$/owner $u_{0}$ blocking $u_{2}$’s future access data $F$. If owner $u_{1}$ is online and willing to do so, go to step \textbf{3}, otherwise, go to step \textbf{4}.

\item \smallskip\textbf{Step 3:} If owner $u_{1}$ is online and willing to do dynamic ownership management, Pri-CSP generates the new symmetric key $K'$ before generating $REK_{u}'$, and sends it back to Pub-CSP. User $u_{1}$ obtains the updated key $K'$ by running $De(sk_{u_{1}}, REK_{u_{1}}')$ once he receives $dp = \left\{CT, REK_{u_{1}}'\right\}$ from Pub-CSP, and $u_{1}$ re-encrypts data $F$ with new key $K'$ to get the updated ciphertext $CT'$. Then $u_{1}$ re-uploads $CT'$ to Pub-CSP.

\begin{figure}[t]
    \centering
    \includegraphics[width=0.95\linewidth]{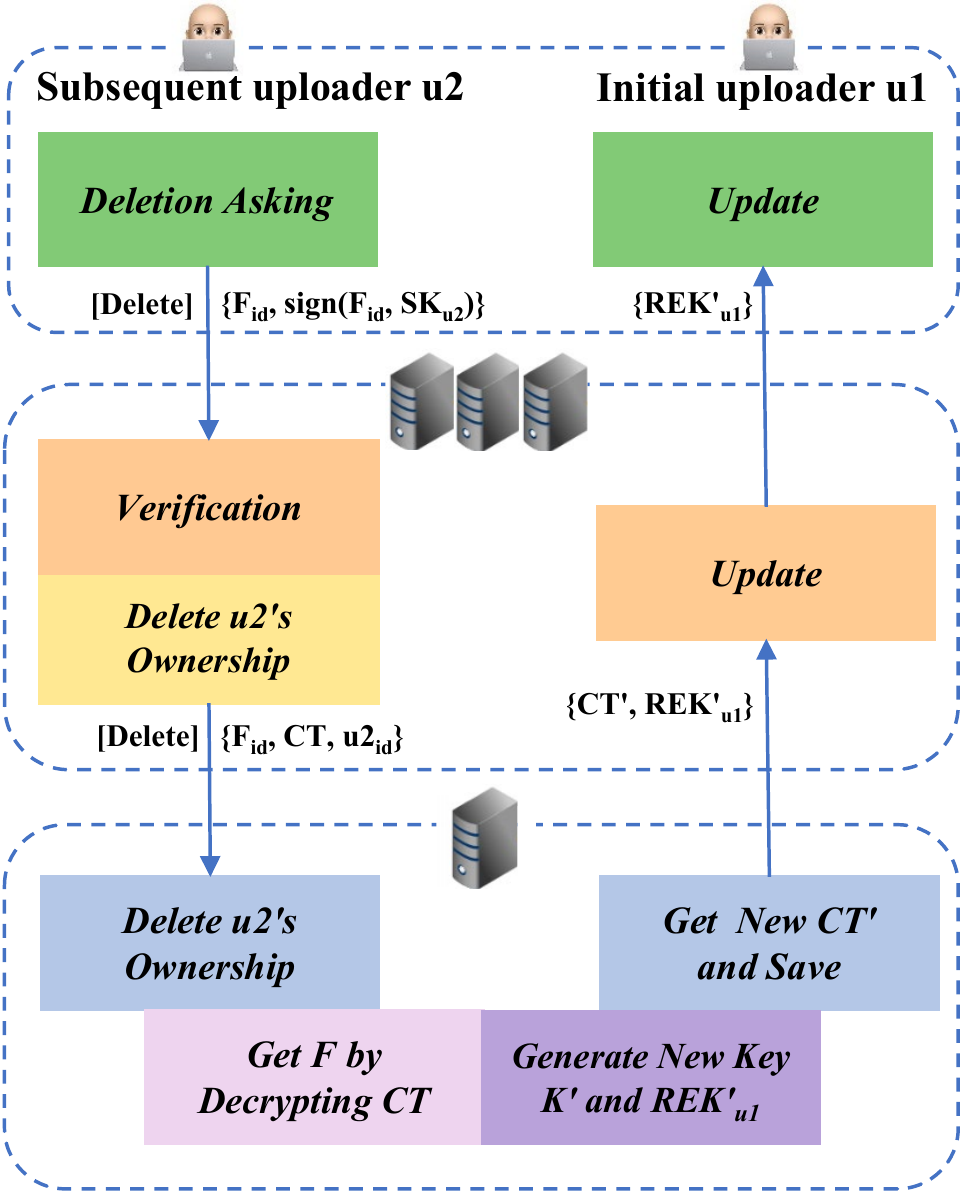}
    \caption{Procedure of ownership revocation of data holder $u_{2}$ when owner $u_{1}$ is offline.}
\end{figure}

\item \smallskip\textbf{Step 4:} If owner $u_{1}$ is offline or asks owner $u_{0}$ representing $u_{1}$ to do ownership management. Pub-CSP sends $(F_{id}, CT)$ to owner $u_{0}$, and $u_{0}$ decrypts $CT$ by running $De(sk_{u_{0}}, EK)$ first before running $Decrypt(K, CT)$ to get plaintext of data $F$. The rest is almost same as step \textbf{3} that $u_{0}$ generates the new symmetric key $K'$ and gets the updated $REK_{u}'$ for the rest users who still own data $F$, and re-encrypts data $F$ with $K'$ before re-uploading the updated ciphertext $CT'$ to Pub-CSP. 

\item \smallskip\textbf{Step 5:} Update the ownership lists saved on the cloud. To make it easier to understand, we assume that there are only two owners $u_{1}$ and $u_{2}$ sharing the same data $F$. The $p_{1}$ and $p_{2}$ are updated as $P_1'$ and $P_2'$ separately.
\begin{equation}
\scriptsize
\nonumber
\begin{aligned}
    &P_1 =\{F_{id}, CT,H(F), (u_{1,id}, REK_{u_{1}}), (u_{2,id}, REK_{u_{2}})\}  \\
    &P_1' = \left\{F_{id}, CT', H(F), (u_{1,id}, REK_{u_{1}}')\right\} \\
    &P_2 = \{F_{id}, X', HC'(F), EK, (u_{1,id}, rk_{u_{0}\to u_{1}}), (u_{2,id}, rk_{u_{0}\to u_{2}})\} \\
    &P_2' = \left\{F_{id}, X', HC'(F), EK', (u_{1,id}, rk_{u_{0}\to u_{1}})\right\} 
\end{aligned}
\end{equation}
\end{itemize}

\noindent\textbf{Ownership Revocation of Data Owner}. The data owner’s ownership revocation basically is the same as the data holder’s except for some details shown as follows.

\begin{itemize}
    
\item \smallskip\textbf{Step 1:} User $u_{1}$ sends a request of data deletion to Pub-CSP by providing  $dp = \left\{F_{id}, sign(F_{id}, SK_{u_{1}})\right\}$.

\item \smallskip\textbf{Step 2:} Pub-CSP deletes the storage record of $u_{1}$ after asking owner $u_{1}$ to pick one from the holders to be the new owner $u_{1}'$, or $u_{1}$ can just pick owner $u_{0}$ as the new owner $u_{1}'$ to manage the dynamic ownership afterwards. Since we have the new owner $u_{1}'$ now, the remaining steps are the exactly same as \textbf{Ownership Revocation of Data Holder}.

\end{itemize}

\section{PERFORMANCE EVALUATION}
\subsection{Comparison with Existing Work}
\textbf{Table 2} is a comparison among four data deduplication schemes, the convergent encryption(CE) \cite{2002Reclaiming}, the randomized convergent encryption (RCE) \cite{2013Message}, the data deduplication with dynamic user management \cite{2018DedupDUM} (DedupDUM) and our scheme. On the basis of encrypted data deduplication, tag consistency, access control, dynamic ownership management, and possession proof.

All the schemes guarantee data confidentiality and privacy by saving encrypted data. However, scheme CE is vulnerable to the tag consistency attack. While other schemes can guarantee data integrity by enabling DU to check the tag consistency of the received data. Scheme DedupDUM solves the dynamic ownership management problem by using the group key generated by DU’s public key, which supports ownership revocation and new user joining. However, they do not verify that the holders hold the entire original file instead of only having a tag, the fake ciphertext and ID, and they also do not consider collusion attacks caused by the dishonest cloud server and attackers.

Different from the previous schemes, our scheme achieves dynamic ownership management by maintaining the ownership list for each data $F$ at Pub-CSP and Pri-CSP separately. Since the re-encrypted key $REK_{u_{i}}$ is generated with $pk_{u_{i}}$, our scheme supports the cloud user revocation and new user joining. Moreover, deduplication is performed after checking that $DU$'s access to file $F$ is unauthorized and he/she does own the whole file. Therefore, the communication overhead can be greatly reduced.

\begin{table}[t]
    \centering
    \caption{Comparison of secure deduplication schemes.}
    \setlength{\tabcolsep}{4.5pt}
    \begin{tabular}{@{}l c c c c@{}}
        \toprule
        Scheme &  CE    & RCE  & DedupDUM & Our Scheme \\ \midrule 
        Encrypted data deduplication & $\surd$ & $\surd$ & $\surd$ & $\surd$\\
        Tag consistency & $\times$ & $\surd$ & $\surd$ & $\surd$\\
        Access control & $\times$ & $\times$ & $\surd$ & $\surd$\\
        Dynamic ownership management & $\times$ & $\times$ & $\surd$ & $\surd$ \\ 
        Possession proof  & $\times$ & $\times$ & $\times$ & $\surd$ \\
        \bottomrule
    \end{tabular}
\end{table}

\begin{table*}[t]
    \centering
    \caption{Communication overhead.}
    \begin{tabular}{@{}l l l l l l@{}}
        \toprule
        \multirow{2}{*}[-3pt]{Scheme} &  \multicolumn{4}{l}{For initial uploader} &  For subsequent uploader \\ \cmidrule(r){2-5} \cmidrule(){6-6} & Upload message size & Download message size & Rekeying message size & Key size & Upload message size \\ \midrule 
        CE &  $C_{C}+C_H+C_{ID}$ & $C_{C}$ & ----- & $C_{K}$ & $C_{C}+C_H+C_{ID}$\\
        RCE & $C_{C}+C_K+C_H+C_{ID}$ & $C_{C}+C_K+C_H$  & ----- & $C_K$ & $C_{C}+C_K+C_H+C_{ID}$\\
        DedupDUM & $C_{C}+C_K+C_H+C_{ID}+C_P$ & $C_{C}+C_K+C_H$ & $C_P$ & $C_K+C_P$ & $C_{C}+C_K+C_H+C_{ID}+C_P$ \\
        Our scheme & $C_{C}+C_H+C_{HC}+C_{ID}$  & $C_{C}+C_K+C_H$  & $C_K$  & $C_K$  & \bm{$C_H+C_{ID}$} \\ 
        \bottomrule
    \end{tabular}
    \vspace{-10pt}
\end{table*}

\subsection{Efficiency Analysis}
The comparison on the basis of communication overhead among four schemes is shown in \textbf{Table 3}. $C_{C}$ denotes the size of the encrypted data, $C_{ID}$ denotes the size of a cloud user’s $u_{id}$, $C_{H}$ denotes the size of a hash code, $C_{HC}$ denotes the size of hash code set of data $F$, $C_{K}$ denotes the size of a key, $C_{p}$ denotes the size of a public key.

For the first upload of data $F$, scheme CE, RCE, and DedupDUM have the same upload message sizes. In our scheme, it increases the size of the hash code set $HC(F)$ used for $DU's$ ownership verification before deduplication. However, for the subsequent upload of $F$, our scheme only uploads $H(F)$ before ownership is verified or access is checked while the other schemes need to re-upload all messages every time as shown in Table 3.

Concerning the rekeying message size, the DedupDUM and our scheme increase the size of the re-encryption key while scheme CE and scheme RCE do not consider key updating. However, even though the group key is used to manage ownership revocation, the encryption key $K$ determined by data $F$ still never changes in DedupDUM once it is settled, which is unsecured since the withdrawn owners can collude with Pri-CSP for getting ciphertext before rekeying. Our scheme updated $K$ as long as there is ownership revocation, which is more secure and computation overhead is accepted since it can be performed by Pri-CSP instead of DU.

\begin{figure}[t]
    \centering
    \includegraphics[width=1\linewidth]{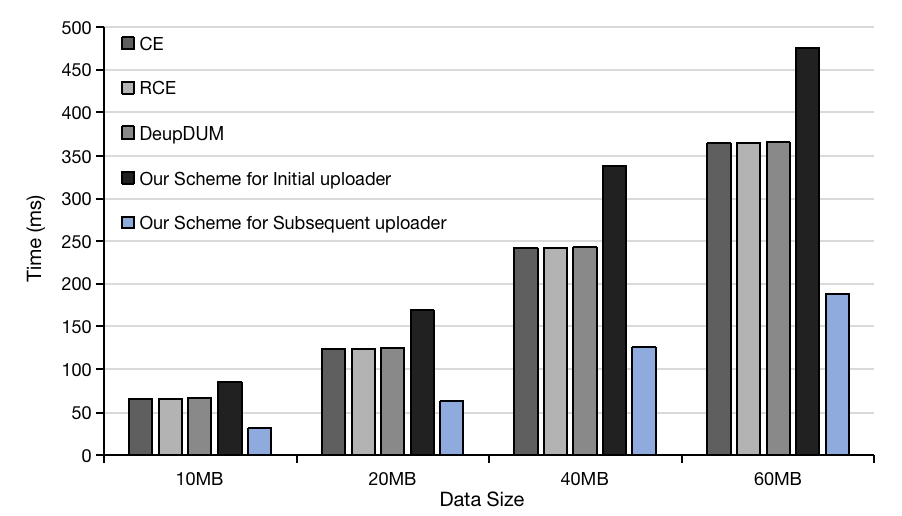}
    \caption{Computation time for upload.}
\end{figure}

\subsection{Performance Evaluation}
In this section, we compare our scheme with previous works.
We follow common practice \cite{2002Reclaiming,2013Message,2018DedupDUM} for fair comparison, each cryptographic operation is implemented by using the umbral library ver. 0.3.0 and Crypto library ver. 1.4.1. We perform data encryption and decryption algorithm with AES where the key is 128-bit. The data size ranges from 10MB to 60MB. The testing environment is Intel(R) Core(TM) i5-7300HQ CPU 3.1 GHz 16.0GB RAM.

\subsubsection{Computation time for upload} 
$\\$We tested the efficiency of the file uploading process under different schemes, and scheme DedupDUM requires the same computations as scheme CE and scheme RCE. As shown in \textbf{Figure 5}, there is a slight increase in our scheme compared with others since the process in our scheme includes calculating hash code and hash code set of data $F$, signing and verifying the signature, decrypting re-encrypted key, and encrypting data $F$ with AES. Moreover, We can see that our scheme has great advantages in the duplicated file uploading process. Since our scheme only uploads $H(F)$ before ownership is verified or access is checked while the other schemes need to re-upload all message every time. The abundant communication cost can be saved in our scheme.

\subsubsection{Computation time for download} 
$\\$Compared with other schemes, we apply the ownership verification mechanism to verify whether DU has access to whole data or not by challenging DU with random $H(F_{x})$ in the hash code set, which can save abundant communication cost when DU does not have ownership. The comparison of computation time for downloading ciphertext among four schemes is depicted in \textbf{Figure 6}.

\subsubsection{Computation time for encryption and decryption} 
$\\$In the encryption stage, the DedupDUM and our scheme resolve the dynamic ownership management problem by performing re-encryption. While our schemes take a shorter time than DedupDUM since Pub-CSP needs to decrypt and re-encrypt the ciphertext for each subsequent uploader in DedupDUM, which will result in high computation complexity as the number of holders grows. The detailed encryption and decryption time for different data sizes (ranging from 10MB to 60MB) is shown in \textbf{Figure 7}.

\begin{figure}[t]
    \centering
    \includegraphics[width=1\linewidth]{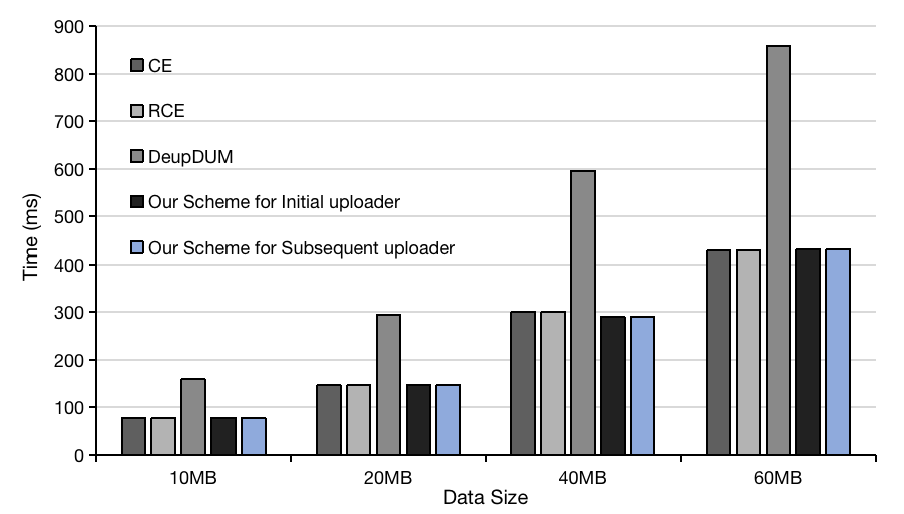}
    \caption{Computation time for download.}
\end{figure}

\begin{figure*}[t]
    \centering
    \includegraphics[width=1\linewidth]{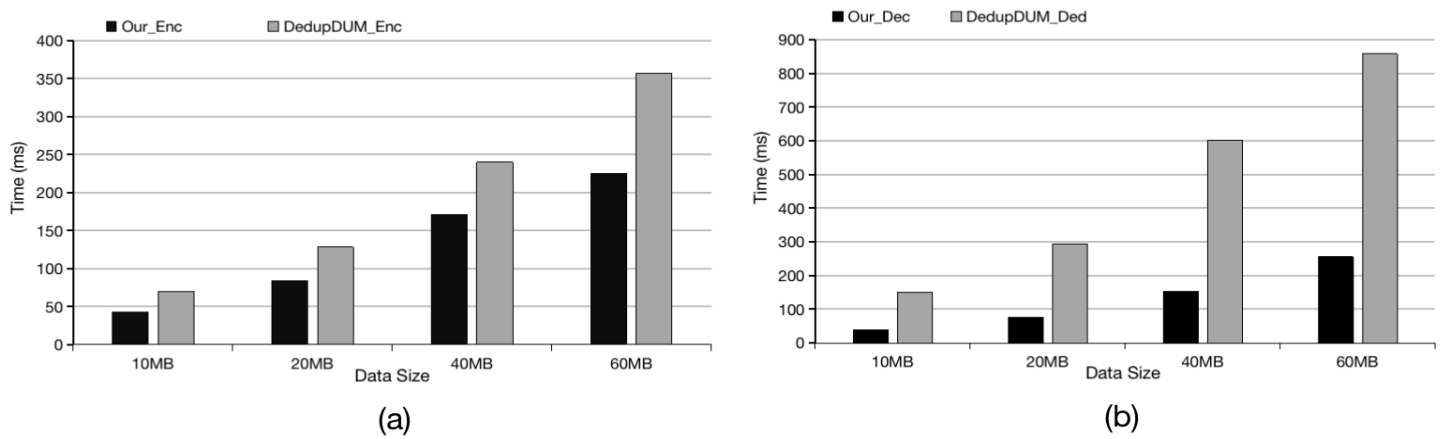}
    \caption{Computation time for (a) encryption and (b) decryption.}
\end{figure*}

\subsubsection{Computation time for deduplication} 
$\\$In this process, the total number of DU is set up as 25, and each of them have a different 2MB-sized file, which means the total size of data saved in the cloud is 50MB when the duplicate radio(DR) is 0. Specifically, when DR is 20$\%$, which means 5 DUs own the same file, and the rest 20 DUs own different files. From \textbf{figure 8}, it shows how computation time changes during each process with DR grows from 0 to 100$\%$. It can be observed that such a process is very efficient, taking less than 0.373 seconds for data upload, and 0.251 for deduplication when the ratio is 100$\%$. Therefore the proposed deduplication scheme can greatly reduce data uploading time.

\section{Security Analysis}
In this section, our scheme security is analyze on the basis of data privacy, data consistency, data ownership verification, ownership revocation, and collusion resistance. 

\subsection{Data Privacy}
In terms of data privacy, the raw data should be prevented from Pub-CSP (honest but curious) and unauthorized data users. Therefore, there are normally two kinds of attacks, separately from Pub-CSP and invalid data users. Firstly, as an attack from Pub-CSP is concerned, what is saved on Pub-CSP is the authorized DU's re-encrypted key that is encrypted through PRE by Pri-CSP and only can be decrypted by DU's private key, since Pri-CSP and authorized users will not collude with Pub-CSP considering their profits, it is impossible for the Pub-CSP to get plaintext by cipher key. Second, supposing an unauthorized user $u_{2}$ requests data $F$ with $(F_{id}, u_{1, id})$ ( $u_{1}$ is valid,and using  $u_{2, id}$ will simply not pass the ownership check), Pub-CSP searches the ownership list $P2 = \left\{F_{id}, CT,H(F), (u_{1,id}, REK_{u_{1}})\right\}$, and return $\left\{F_{id}, CT, REK_{u_{1}}\right\}$ to user $u_{2}$ based on $(F_{id}, u_{1,id})$. Since $REK_{u_{1}}$ can only be decrypted by the private key of user $u_{1}$, it is computationally infeasible for user $u_{2}$ to obtain plaintext of $F$ by decrypting $CT$ with $REK_{u_{1}}$. Therefore, data privacy against the honest-but-curious Pub-CSP and unauthorized users is guaranteed.

\begin{figure}[t]
    \centering
    \includegraphics[width=1\linewidth]{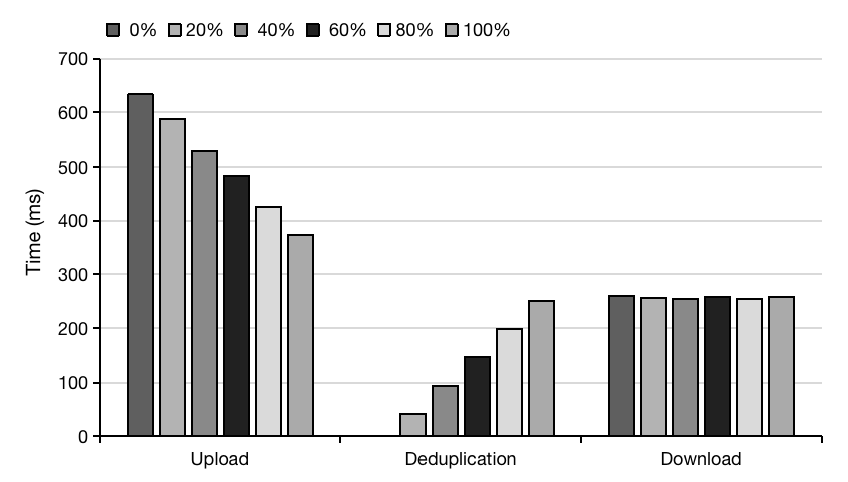}
    \caption{Computation time for each process with a duplicate ratio grows.}
\end{figure}

\subsection{Data Consistency}
In the data deduplication scheme, data integrity may be threatened by a poison attack on tag consistency, which could be identified during the decryption process by the data holders. Assuming that attacker $u_{2}$ owns the same data $F_{A}$ with authorized user $u_{3}$, $u_{2}$ uses $F_{B} \neq F_{A}$ to generate a fake ciphertext $CT_{B}$, then uploads $(H(F_{A}), CT_{B})$ to Pub-CSP to pretend to be $CT_{A}$. When the authorized user $u_{3}$ wants to uplaod $F_{A}$, $u_{3}$ sends $H(F_{A})$ to Pub-CSP to check duplication. Since $H(F_{A})$ exists, Pub-CSP asks Pri-CSP performing dedupliction. Pub-CSP sends $(CT_{B}, REK_{u_{3}})$ back to user $u_{3}$ after dedupliction, and $u_{3}$ checks whether $H(Decrypt(De(sk_{u_{3}}, REK_{u_{3}}), CT_{B})) = H(F_{A})$ holds or not, if this is not consistent then $u_{3}$ drops the messages and reports this to Pub-CSP. Therefore, our scheme guarantees the data integrity.
\subsection{Data Ownership Verification}
In our scheme, the data ownership is verified by challenging DU with random $H(F_{x})$ in hash code set (randomly select specific parts of the data, e.g., the hash code of 10.5-14.3$\%$ of $F$). $F_{x}$ is randomly selected and function $H(*)$ is non-invertible, therefore it is impossible to calculate $H(F_{x})$ without the original plaintext. 
\subsection{Ownership Revocation}
The ownership withdrawn data user should be restricted to access data $F$. Our scheme guarantees ownership revocation by using Pri-CSP playing a role as owner $u_{0}$. Whenever the data owner $u_{1}$ withdraws his/her ownership or some other holders may request to delete or modify their data, the owner $u_{0}$ deletes the ownership information of the requestor from the ownership list, and $u_{0}$ (if $u_{1}$ is offline or revokes his/her ownership or asking $u_{0}$ update for him/her) re-encrypts the plaintext with the new symmetric key before re-uploading it to Pub-CSP, and updates the re-encrypted keys of the rest users. Hence, the withdrawn data owner will not be able to pass the access check and decrypt the latest ciphertext with the non-updated cipher-key. 
\subsection{Collusion Resistance}
Since Pri-CSP is fully trusted, we further discuss the collusion attacks launched by dishonest Pub-CSP and attackers. First, if unauthorized user $u_{1}$ colludes with dishonest Pub-CSP for getting plaintext of data $F$, Pub-CSP will ask Pri-CSP performing deduplication for $u_{1}$ with faking information. Pri-CSP will verify $u_{1}$'s ownership of data $F$ before sharing the re-encrypted key $REK_{u_{1}}$ of $F$ with $u_{1}$. Since $u_{1}$ does not have plaintext, he/she can not pass the ownership check to get cipher-key even though Pub-CSP sends $u_{1}$ the ciphertext dishonestly. Second, since each cipher-key can only be decrypted by the corresponding user, the unauthorized users are not able to decrypt them even if they collude with each other. Therefore, our scheme guarantees collusion resistance.

\section{CONCLUSION}
In this paper, we proposed a secure and practical scheme that managed the encrypted data with deduplication, based on ownership challenge under a hybrid cloud architecture, where Pub-CSP manages the storage and Pri-CSP plays a role as owner $u_{0}$ and proxy at the same time to perform deduplication and dynamic ownership management. Further, our scheme proves that the owner holds the real data alone pass the data ownership, and encrypted data can be securely accessed because only authorized data holders can obtain the symmetric keys used for data decryption. Security analysis, comparison with existing work, and implementation-based performance evaluation show that our scheme is secure and efficient, and resists collusion attacks and duplicate faking attacks.

\section*{ACKNOWLEDGMENT}
This work was supported in part by the National Innovation 2030 Major S\&T Project of China under Grant 2020AAA0104203, and in part by the Nature Science Foundation of China under Grant 62006007. 
We thank all the anonymous reviewers for their constructive comments and suggestions. 
The corresponding author of this paper is Yuesheng Zhu.

\bibliographystyle{IEEEtran}
\bibliography{IPCCC2022}

\end{document}